# Analyzing Computing Undergraduate Majors from Job Market Perspective

**Yazeed Alabdulkarim[1], Khalid Alruwayti[2], Hamad Alsaleh[3], Sultan Alfallaj[4], Ahmed Bablail[5], Abdulaziz Almaslukh[6]**

Department of Information Systems, College of Computer and Information Sciences, King Saud University, Saudi Arabia
*Email: yalabdulkarim@ksu.edu.sa (corresponding author)[1], 441103156@student.ksu.edu.sa[2], haalsaleh@ksu.edu.sa[3], 441105970@student.ksu.edu.sa[4], 441102924@student.ksu.edu.sa[5], aalmaslukh@ksu.edu.sa[6]*

***Abstract***: *The demand for computing education increases due to the rapid development of technology and its involvement in most daily activities. Academic institutes offer a variety of computing majors, such as Computer Engineering (CE), Computer Science (CS), Information Systems (IS), Information Technology (IT), Software Engineering (SE), Cybersecurity (CSEC), and Data Science (DS). Since a major objective of earning a bachelor's degree is to improve career opportunities, it is crucial to understand how the job market perceives these computing majors. This study analyzed the relationships between various computing majors and the job market in Saudi Arabia, using LinkedIn public profile data, discovering insights into the strong relationship between the focus of certain computing majors and the employment of relevant job positions. Moreover, job category trends were analyzed over the past ten years, observing that demands for System Admin and Technical Support positions declined while demands for Business Analysis and Artificial Intelligence & Data Science inclined. This study also compared earned professional certifications between different computing major graduates that correspond to job position findings.*

## I. INTRODUCTION

Technology is essential for most daily activities. People use mobile applications and computers to work, communicate, have fun, shop, etc. Global spending on digital transformation is expected to reach USD 3.4 trillion in 2026 [1]. E-commerce sales are expected to exceed USD 8 trillion in 2026, with 24% of retail purchases placed online [2]. Consequently, more people are interested in studying computing and technology fields, and the demand for new technologies is increasing, creating a new array of jobs that need to be occupied. These factors produce a high demand for academic computing degrees. In academia, the term computing is often used to capture all disciplines that focus on the scientific study of computers. Academic institutions offer programs in various computing disciplines and majors to attract graduate-qualified applicants and students. For example, Saudi Arabian universities offer various computing programs to meet recent demands in computing specialties [3]. The Association for Computing Machinery (ACM), IEEE Computer Society (IEEE-CS), and others list seven computing subdisciplines or majors for undergraduate degree programs in their 2020 CCR [4]. These subdisciplines are Computer Engineering (CE), Computer Science (CS), Information Systems (IS), Information Technology (IT), Software Engineering (SE), Cybersecurity (CSEC), and Data Science (DS). These programs focus on different areas. For example, CS has a strong theoretical emphasis, while IS focuses on information and understanding of business aspects to enable transformative organizational changes.

Students in a computing program are expected to have better skills in their area of focus. However, it is unclear how these academic differences are reflected in the job market. For example, are IS graduates preferred for specific job positions? The abundance of career opportunities after graduation is one of the main reasons for selecting a bachelor's major [5]. With more than 50% of students changing their major at least once [6-8], understanding the differences between computing majors from the job market's perspective should help students make better-informed decisions about their chosen study program. A limitation of previous studies is that they did not analyze post-graduation job outcomes and did not take into account recent advancements in the technological sector, which have a direct effect on majors and career choices. Furthermore, there is a lack of periodic examinations of specific specialties and their popularity among students. Additionally, little research has been done on the connection between academic programs and job selection at widely used job placement sites, particularly in Saudi Arabia, which has a rapidly developing technological infrastructure.

This study focused on understanding how various bachelor's degree computing majors are perceived from the job market point of view, considering new trends in job positions, by collecting and analyzing LinkedIn profiles to answer the following research questions:

- RQ1: What are the differences between the computing majors for fresh graduate job positions?
- RQ2: What are the new trends in job positions for fresh computing graduates?
- RQ3: What are the top-earned professional certificates for various computing graduates?

This study provides insight to help prospective students and computing program coordinators make better decisions. Students should select the computing subdiscipline, specialty, or major that addresses their job market concerns. For example, a student interested in a specific job may select the computing major with more hiring opportunities in that area of interest. On the other hand, program coordinators can adjust their curriculums based on given job market insights. For example, program directors can add courses in response to new trends in the job market to appeal to new students.

## II. RELATED WORKS

There is a lack of research on students' major choices and job progression after graduation. In [9], LinkedIn profiles of computing program alumni from a medium-sized university in the southeast of the United States were examined to address questions about accountability in higher education. This study aimed to provide a comprehensive overview of career progression, however, it only focused on a closed group of graduates from one mid-size university, which may not represent all graduates from mid-sized universities. Other studies analyzed social media sites to gain insight into working adults. In [10] the influences of personality, emotion, interest, and education on career advancement were discussed. Tweets and job information were collected from 150 thousand user profiles, using Twitter and LinkedIn data, extracting 29 traits for each user, including thinking styles, interests, and personalities. Similarly, in [11], it was examined how employees and their co-workers interact on Facebook and how satisfied they were with their last jobs.

The studies in [12-13] used a cross-sectional mixed-method design to examine how undergraduate Computer Science (CS) students choose their majors and how these choices related to their expected careers after graduation. Their method may influence students' answers by causing them to provide answers that they perceive to be socially desirable, potentially affecting the accuracy and quality of their responses. Furthermore, these studies lack longitudinal data, as they tracked participants' development only before graduation. Other studies used machine learning models to predict employment-related issues. In [14], a collaborative framework was built using machine learning classifiers to predict and bridge the gap between graduates' skills and employer expectations, achieving 89% F1-Score accuracy for identifying relevant features for employability. Similarly, in [15], CareerRec was introduced, which is a recommendation system that uses machine learning algorithms to help IT graduates choose a suitable career path based on their skills. The system was trained and tested on a dataset of 2255 IT sector employees in Saudi Arabia, and the experiments showed that the XGBoost algorithm outperformed other models, achieving the highest accuracy (70.47%) in predicting the best-suited career path among the three classes.

Several limitations have been identified in previous studies. At first, most studies interviewed students or analyzed their choices before graduation, but did not analyze what jobs they held afterward. In general, studies that analyzed students' progression after graduation analyzed students who graduated from specific institutions. Secondly, in light of the current popularity of computing majors, particularly with the advent of new technologies such as artificial intelligence and generative models such as ChatGPT [16], it is necessary to examine whether specific specialties are being preferred by students over the last few years. For this reason, it is necessary to perform a recent analysis regarding job selection and majors. Finally, in the Saudi Arabian context, no recent study has examined the connection between major choices and first job selections on a widely used job placement site such as LinkedIn. As a result, it is critical to analyze this context. This is in particular the result of government initiatives and transformation of the technological infrastructure and the popularity of LinkedIn for employment in the region [17].

## III. COMPUTING DISCIPLINE AND MAJORS

According to ACM and IEEE-CS, the word computing refers to "a goal-oriented activity requiring, benefiting from, or associated with the creation and use of computers" [4]. In general, a computing specialty or major may refer to a specific area of expertise that requires specialized knowledge, skills, and experience within the field of computing. The ACM has divided the discipline of computing into seven defined subdisciplines or majors:

- Computer Engineering (CE): CE combines concepts of computing and electrical engineering into a single major. CE focuses on designing, implementing, and constructing hardware components of computers, such as circuitry, hardware interfaces, and software components, such as firmware and network protocols.
- Computer Science (CS): CS has a strong emphasis on programming, algorithms, data structures, and software development. Generally, CS is heavily focused on the theoretical side of computing compared to other computing majors and is often viewed as a fundamental discipline in the computing field.
- Information Systems (IS): IS deals with improving business processes using computing principles, bridging the gap between management and technology. It focuses on building systems capable of transforming data into information by collecting, processing, and analyzing, which are then stored and used in decision-making.
- Information Technology (IT): IT addresses the technological needs of organizations and their business users, focusing on implementing, designing, installing, and maintaining hardware and software solutions, such as programs, computers, networks, and servers, on which the organization relies to achieve its goals effectively.
- Software Engineering (SE): SE combines CS and CE concepts into a major that focuses on the development of software using best practices and approaches, taking into consideration various software characteristics, such as reliability, usability, maintainability, and security. SE focuses on methods to create software.
- Cybersecurity (CSEC): CSEC focuses on protecting and mitigating cyber threats against an organization's computer systems, such as hacking, malware, and cyber espionage. CSEC professionals implement a variety of technical procedures, including encryption, firewalls, and endpoint security solutions, as well as organizational procedures, including policies, risk management, and training.

- Data Science (DS): DS is an interdisciplinary field that combines statistical analysis, CS, and domain expertise to extract insights and knowledge from large amounts of raw data. DS has become increasingly important in recent years due to the ever-expanding economy of data, which created a need for automated processes to analyze the data and generate insights to drive innovation and decision-making in organizations.

## IV. RESEARCH METHODOLOGY

This study collected LinkedIn profile data to answer its research questions empirically.

### *A. Data Collection and Preprocessing*

Search engines were used to find public LinkedIn profiles for the target group, using keywords, such as computing majors and Saudi university names, while limiting results to the LinkedIn website. The focus was placed on common bachelor's degree computing majors in Saudi universities, which are CS, IS, CE, SE, and IT [3]. As a result, 3,654 LinkedIn profiles were collected and web scraping tools [18-20] were used to extract and preprocess these profiles. Then, duplicates, graduates with irrelevant bachelor's degrees and profiles missing critical data, such as graduation year and employment history were removed. Profiles that contained inaccurate or missing first-position information were also removed, as these data were essential for this study.

After preprocessing, 2,055 profiles were kept, representing individuals with bachelor's computing degrees who reside in Saudi Arabia. Data on education, professional certification, and work experience were obtained for each profile, as shown in Figure 1. Additionally, the institution name, computing major, and earned academic degree dates were fetched, as well as the certificate titles and provider names for professional certifications. For work experience, the job position, company name, and start and end dates were collected. Focus was placed on the first job after graduation as it is more likely to be influenced by the bachelor's degree than the following jobs. The effect of the bachelor's degree is less in future job positions as other factors, such as work experience, become part of the employment decision. The first job positions were classified into broad areas of interest, as shown in Table I. These keywords were used for classification and each job was manually inspected to resolve problems and assign multiclass profiles to the most appropriate class. These job categories provide essential insight into jobs rather than being distracted by fine-grained job positions. This is particularly true in this dataset, which contains 1,069 unique job positions in 2,055 records.

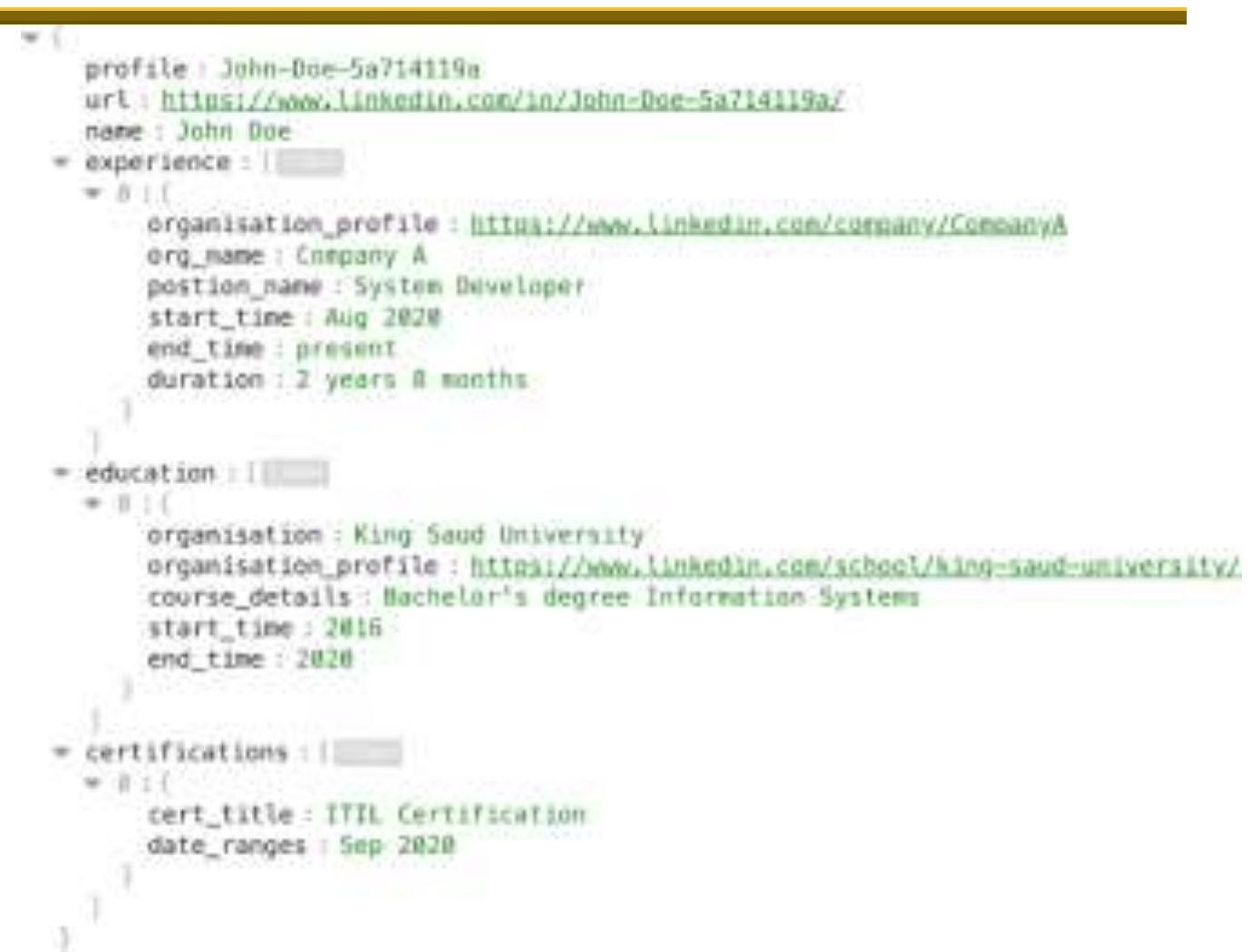

Fig. 1. Extracted LinkedIn profile data.

TABLE I. CATEGORIES OF JOB POSITIONS

| Job Position Category | Keywords |
|---|---|
| Development | Developer, software, development, backend, back end, programmer, implementer, full stack |
| Business Analysis | kw |
| Artificial Intelligence & Data Science (AI & DS) | AI, data, rpa, automation, machine learning, deep learning, computer vision, intelligence, robot, robotic |
| Technical Support | Support, help desk, call center, technician |
| Quality Assurance | Quality, assurance, testing, tester, qc |
| Project Management | Project, product manager, product owner |
| Network | Network, noc, lan, communication |
| Enterprise Systems | Sap, erp, oracle, integration, enterprise, siebel |
| System Admin | Infrastructure, dba, database, operations, admin, cloud, administrator |
| Security | Security, cyber, cybersecurity, threat, soc, risk, secure, defence, forensics, dfir, vulnerability, grc, governance, compliance |
| Sales | Sales, sale |
| Graduate Development Programs (GDP) | Program, graduate, talent, gdp |
| Education/Research | Professor, researcher, trainer, dean, research |

### *B. Data Description*

Table II shows the features collected after preprocessing the data. Figure 2(a,b) shows that the 2,055 profiles in the collected dataset consisted of approximately 21% of CS, 20% of SE, 18% of IT, 18% of IS, and 16% of CE graduates, who graduated from King Saud University (64%), King Abdulaziz University (20%), and King Fahd University of Petroleum and Minerals (16%). Figure 2(c) shows a breakdown of bachelor's degree years in three-year inclusive intervals. The dataset had recent graduates, with about 44% graduating between 2023 and 2018.

(a)

(b)

(c)

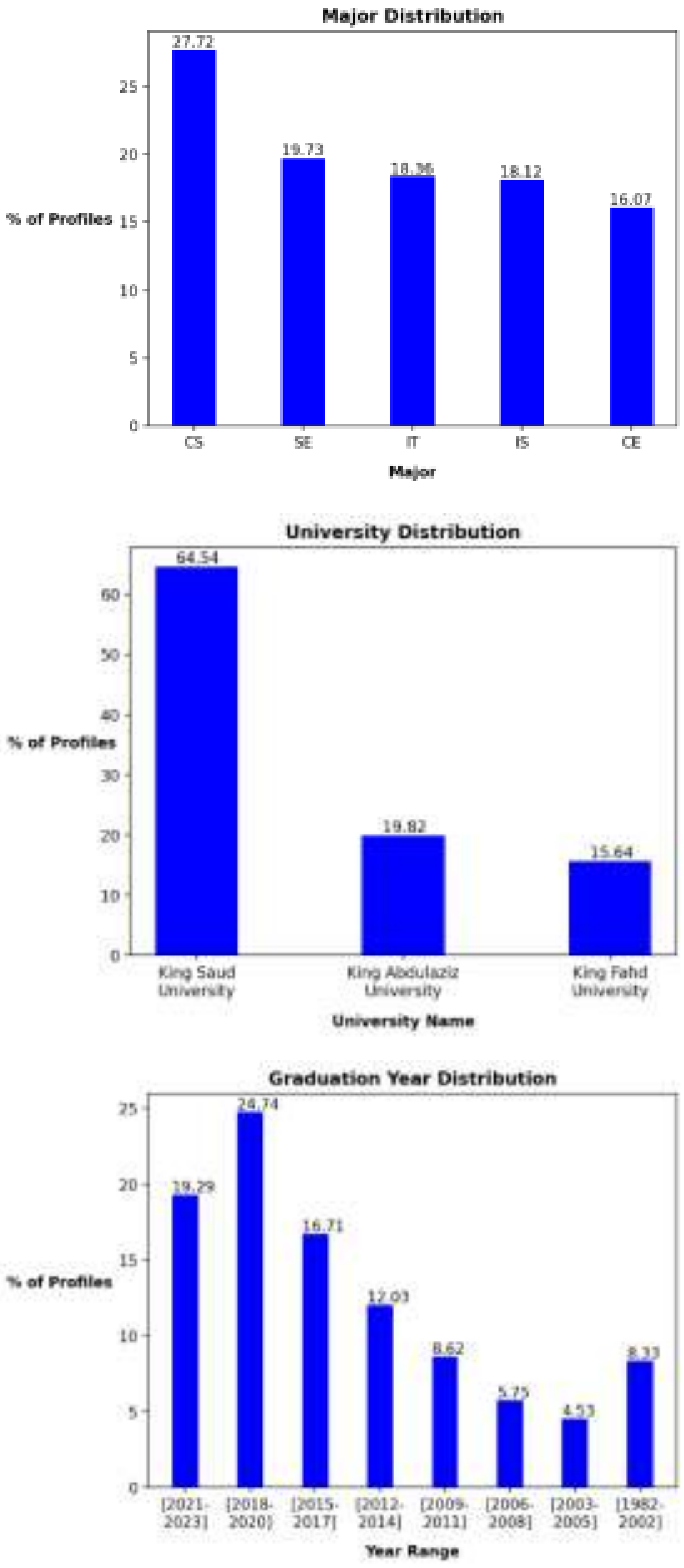

Fig. 2. Statics of collected profiles: (a) Bachelor's degree computing majors, (b) Graduating universities, and (c) Graduation year breakdown.

TABLE II. LIST OF COMPUTED FEATURES

| Feature | Description |
|---|---|
| Computing major | Bachelor's degree computing major, e.g., CS and IS |
| University | University name of the earned bachelor’s degree |
| Graduation Year | Bachelor's degree graduation year |
| First position | The first job position name after graduation |
| First position category | Category of the first position, as shown in Table I |
| First position year | Start year of the first position |
| Industry type | The industry type of the first joined company |
| Years since bachelor's degree | Number of years passed since earning the bachelor's degree |
| Certifications | List of professional certifications listed in the profile |
| Positions | List of all job positions along with their start and end dates listed in the profile |

## V. RESULTS

The collected profiles along with their features were analyzed to answer the study's research questions.

### A. *RQ1: What are the Differences Between Computing Majors for Fresh-Graduate Job Positions?*

Figure 3 shows the top ten job categories for each major of fresh graduates, where each bar is a job category showing its ratio for specific computing major graduates. For example, Software development positions account for 27% of CS fresh graduates' jobs. This is the most common job category for fresh graduates as it contributes 42% for SE, 27% for CS, 25% for IT, 15% for IS, and 13% for CE. This result is reasonable, as software development includes various entry-level positions suitable for fresh graduates. Notably, software development constitutes the majority of job positions for new graduates across all computing majors except CE and IS. The most common job categories for CE and IS graduates are Business analysis and Network, respectively. The focus of CE on networking topics and IS on business aspects in Saudi universities [3] justifies this result. Business Analysis and System Admin are other top job categories across all computing majors. Business analysis jobs represent 31% of IS, 17% of SE, 11% of IT, 10% of CS, and 6% of CE, while System Admin jobs are common in CE (15%) and IT (13%).

Figure 4 shows the industry type of fresh graduate employers, based on LinkedIn's classification. The IT Services industry employs the most fresh computing graduates (21%), followed by Telecommunications (12%) and Banking (8%) industries. These top three business categories account for 41% of fresh graduate hiring.

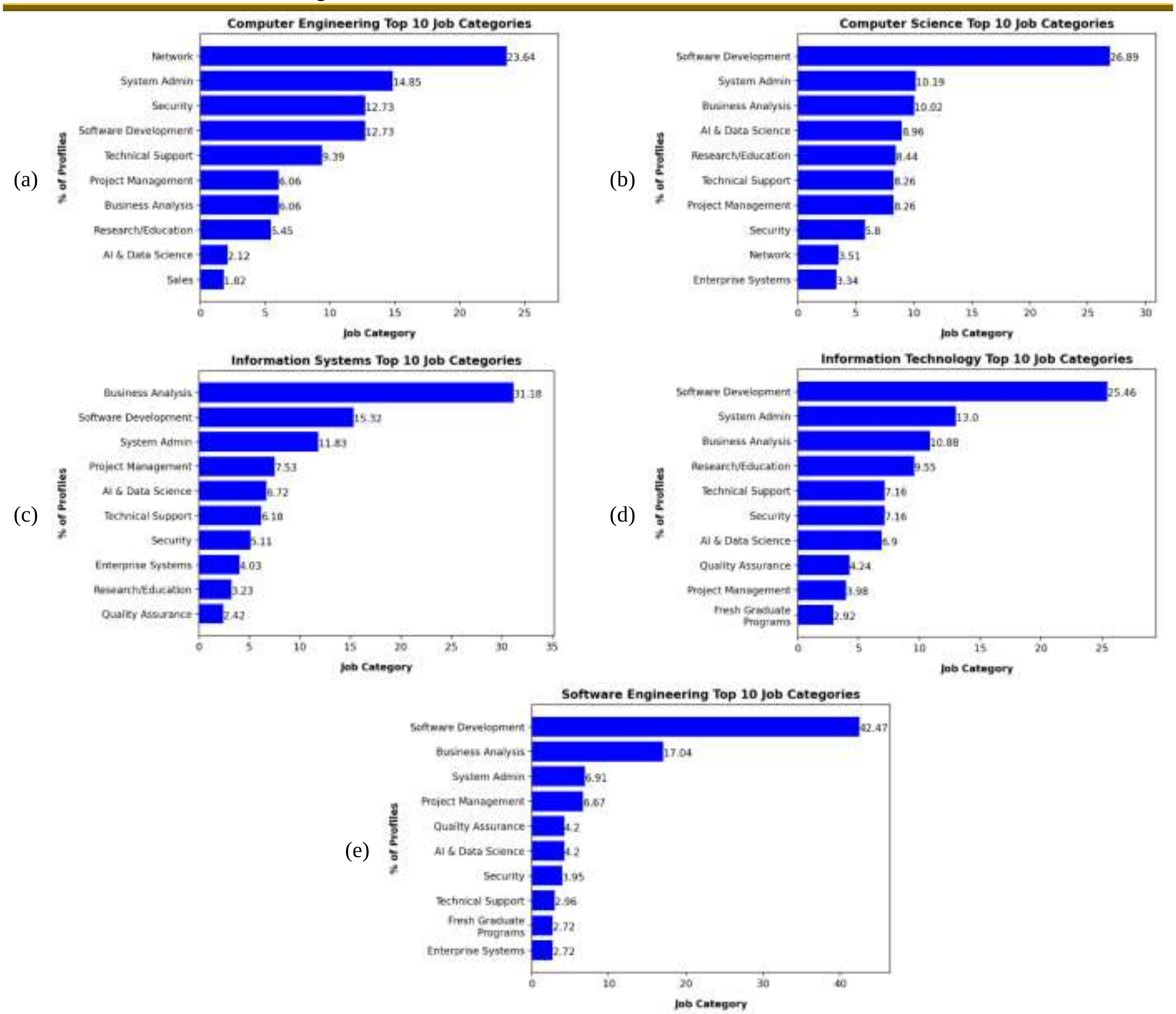

Fig. 3. Top 10 job categories for each computing major: (a) CE, (b) CS, (c) IS, (d) IT, and (e) SE.

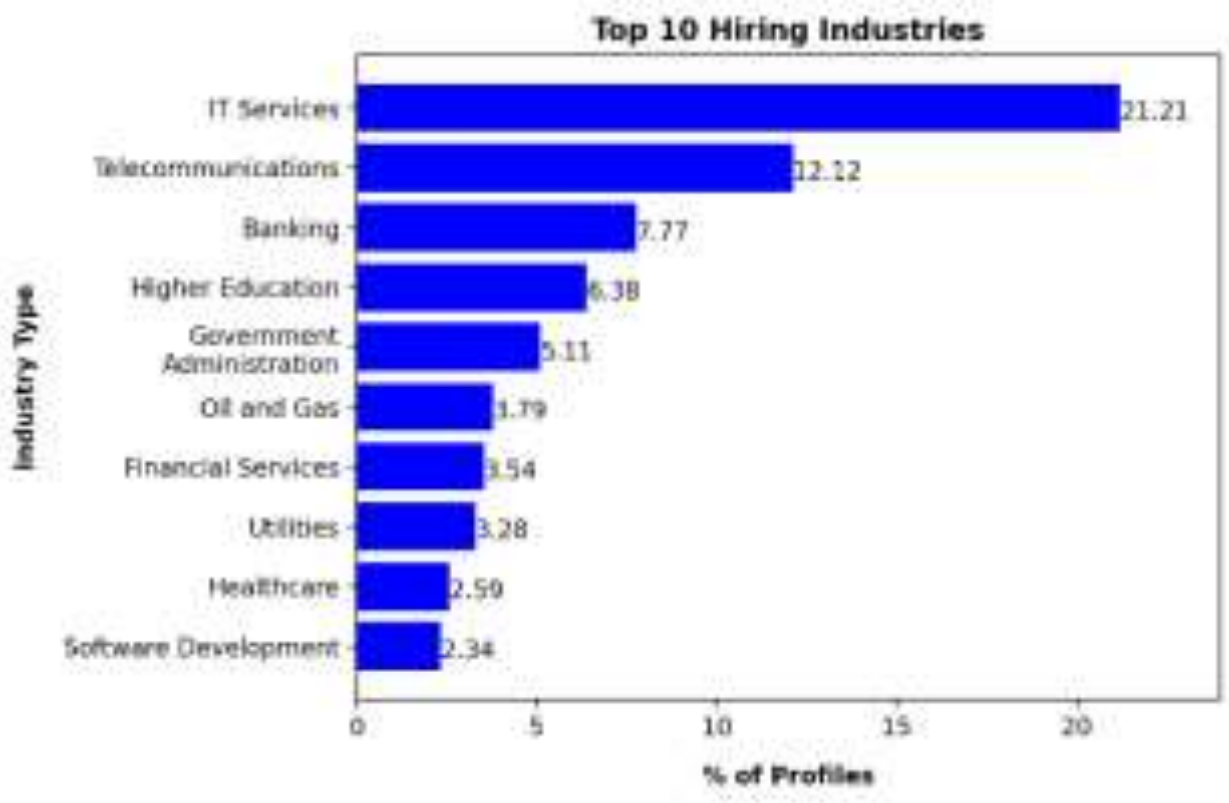

Fig. 4. Top 10 hiring industries for computing graduates.

Another essential aspect for fresh graduates is the time to find a job. The number of required years for new graduates to find a job was calculated. About 90% of computing graduates found jobs within two years of graduation. Most of the profiles only recorded the year of graduation on first hiring, which prevented computing the time to find a job in months rather than years.

Figure 5 shows the average number of years to reach certain job positions. The data indicated that it takes an average of three years to become a Senior Engineer, IT Project Manager, or Senior Database Administrator. This number increases to 9 and 11 years for IT Consultant and IT Director, respectively. For the top of the pyramid, it takes an average of 18 and 24 years to be a Chief Information Officer and Chief Executive Officer, respectively.

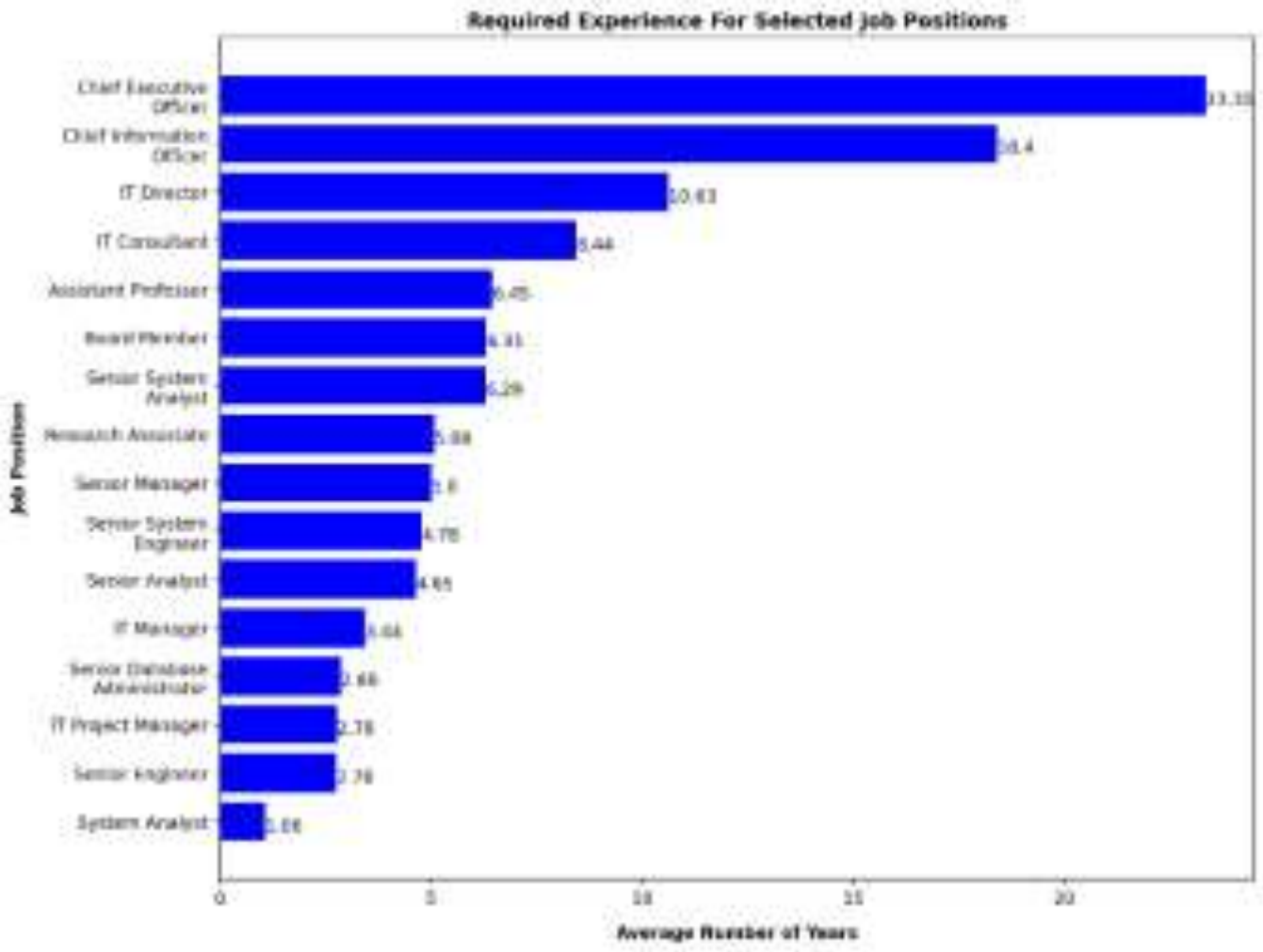

Fig. 5. Average number of experience years to reach selected positions.

## B. *RQ2: What are the New Trends in Job Positions for Fresh Computing Graduates?*

For this question, it was necessary to look back at historical job positions for fresh graduates to identify new trends in employment. The collected profiles were clustered into 2-year intervals based on their graduation year, creating five clusters. Each cluster was associated with an interval of start and end dates. An interval of [*x*-*y*] includes profiles with bachelor's degree graduation years greater than or equal to *x* and less than or equal to *y*. Figure 6 shows the six most popular job categories and their trends across these year-range clusters. The x-axis shows the graduation year intervals, and the y-axis shows the percentage of profiles within a cluster holding a specific job category. Software Development is the highest job category across all years. Business Analysis, Artificial Intelligence & Data Science, and Security job positions have increased in recent years, while System Admin and Technical Support jobs are declining. Artificial Intelligence & Data Science jobs started to grow eight years ago steadily. Notably, Business Analysis jobs incurred an ascending movement while Software Development jobs descended four years ago.

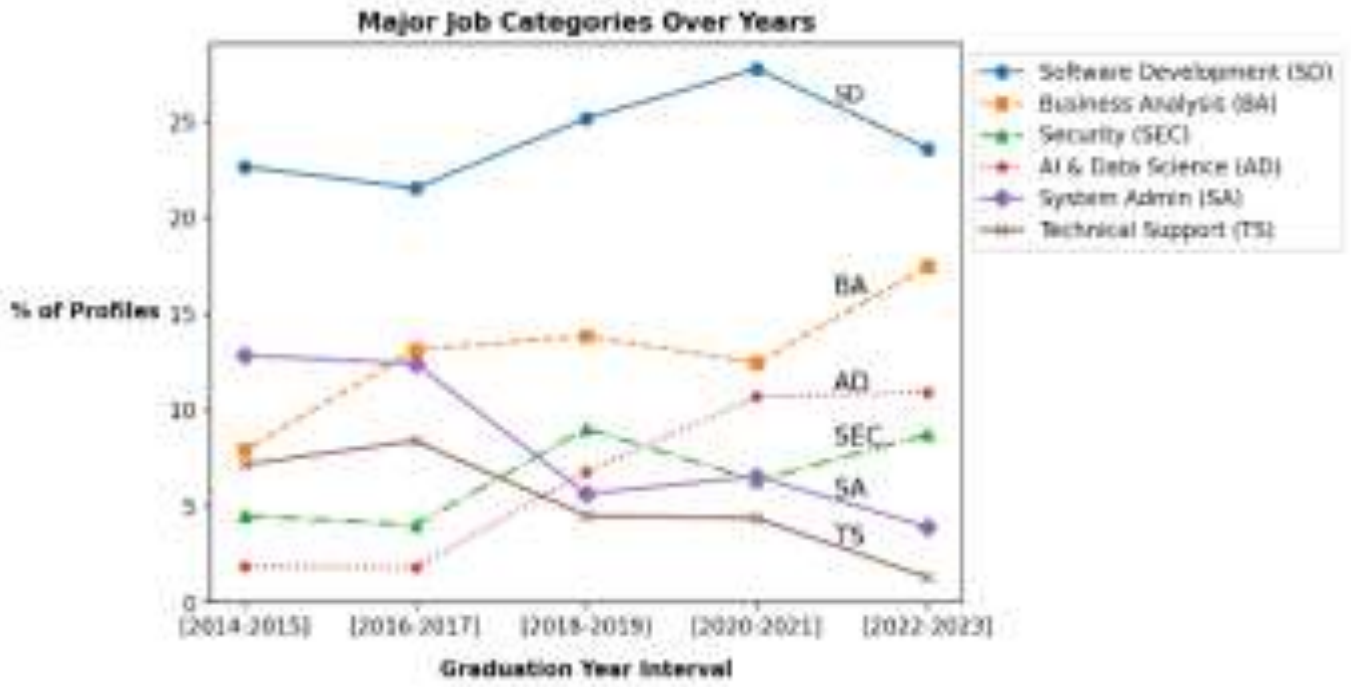

Fig. 6. Job category trends over the past 10 years.

## C. *RQ3: What are the Top-Earned Professional Certificates for Various Computing Specialties?*

Figure 7 shows the top eight earned certificates across all computing majors. The Information Technology Infrastructure Library (ITIL) is the highest earned certification for all computing majors. ITIL is an organizational framework for IT professionals, making it a desirable option for all graduates to enhance career opportunities.

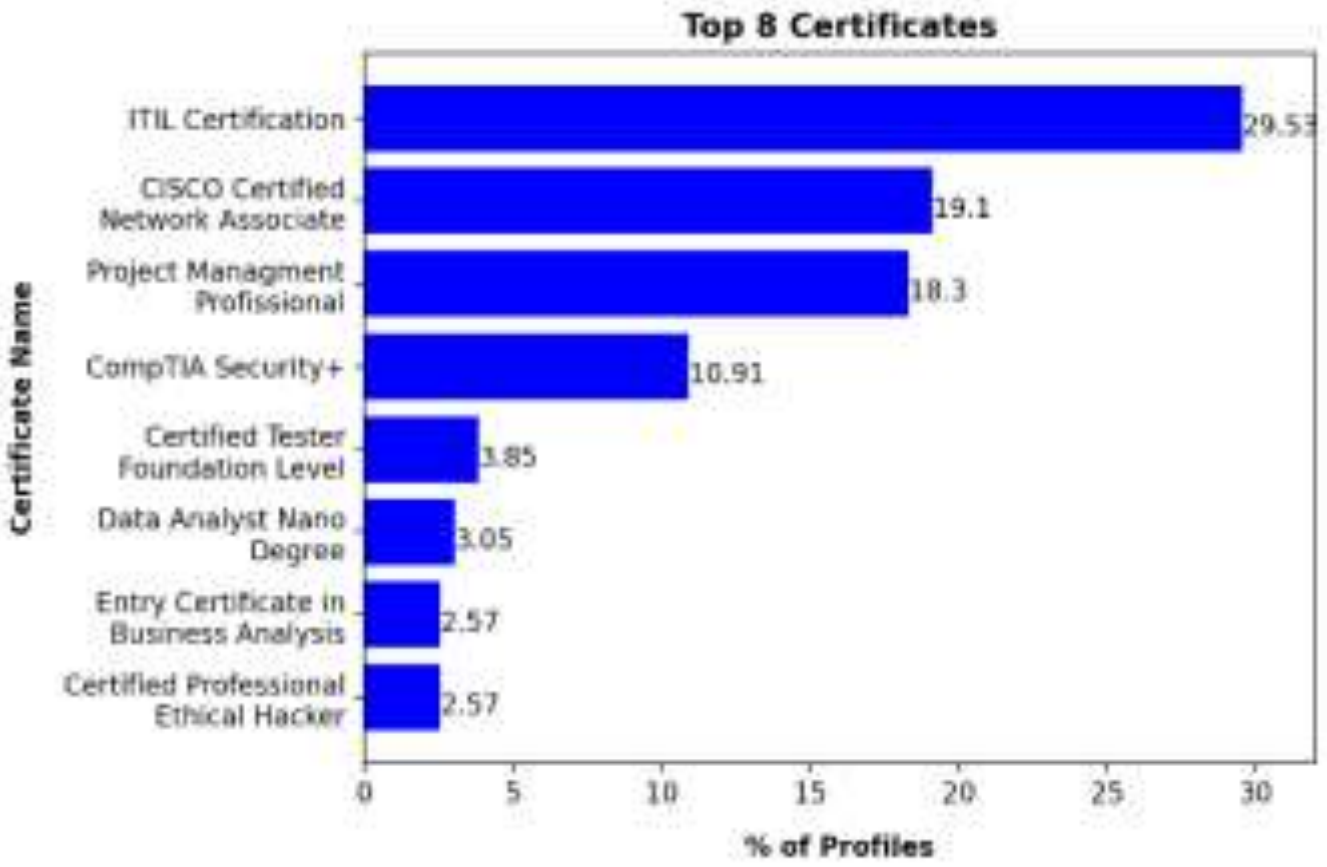

Fig. 7. Top 8 professional certificates earned by computing graduates.

Figure 8 shows that the CISCO Certified Network Associate (CCNA) was the second highest-earned certificate, especially for CE graduates. CCNA certification is important for employees in computer networking positions. The preference of CE graduates for CCNA agrees with the results of RQ1, that computer network positions are common for CE. Another common certification for all computing majors is Project Management Professional (PMP). This result is reasonable, as project management positions were present in all majors.

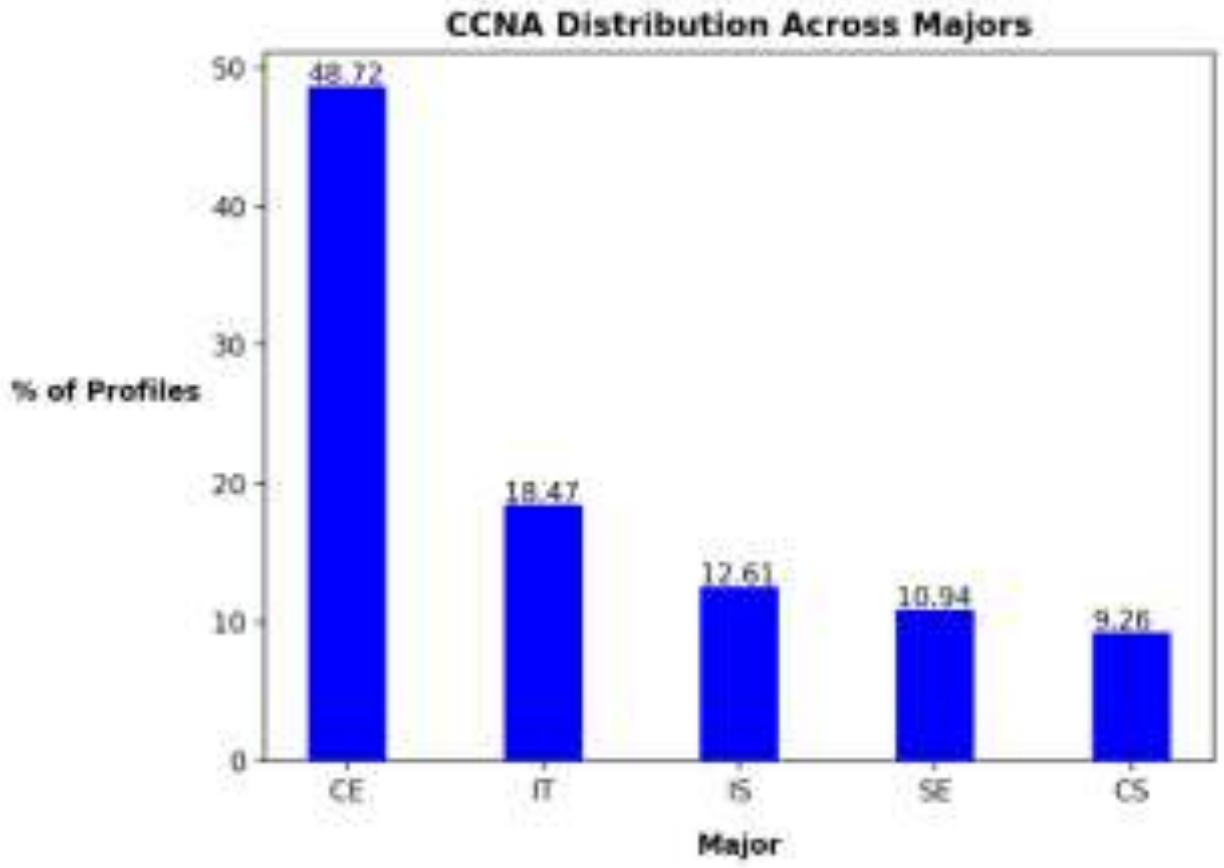

Fig. 8. Earned CISCO Certified Network Associate (CCNA) certificates breakdown across computing majors.

## VI. DISCUSSION

This analysis has rich insights that could help newly enrolled students and education leaders better understand the different computing fields and the expected job market opportunities for fresh graduates. First, the results show a strong relation between the focus of a computing major and the hiring of relevant job positions. For example, graduates of IS major, which includes emphasis on business aspects, are more frequently hired as business analysts. Similarly, CE graduates are more likely to be hired as network engineers. This job-major relationship may be justified by the strong early preparation of students for these relevant jobs. In response to this insight, academic program directors should design a more focused curriculum for coursework, capstone project, and cooperative training to emphasize these roles for each major. Additionally, raising students' awareness of their chosen computing major's concentration is an essential intervention as it may have a significant impact on their future career opportunities. Second, the study finds that emerging roles, such as Artificial Intelligence and Data Science roles, are increasing in demand while other roles, such as Technical Support, are declining, as these roles have been partially automated [21]. Thus, academic programs should be adaptive to technology trends by introducing new courses or tracks that focus on meeting job market demands. Last, professional certifications could have an impact on fresh graduates' career growth. Popular top-notch certificates, such as ITIL, could be included as extracurricular courses for all majors, while specialized certificates, such as CCNA, could be included in relevant majors, for example, CCNA in CE as an extracurricular course. These certificates could benefit students in accelerating their career development.

Although the analysis of the dataset reveals interesting insights into different computing fields, the collected dataset and the study had specific limitations. First, the dataset was collected from public LinkedIn profiles. Therefore, the profile information collected may not be factual. Verifying the accuracy of a single profile is a difficult task that involves significant human effort and expensive techniques. Thus, verifying 2,055 profiles is even more challenging and time-consuming. Second, the profiles that graduated in 2015 and later are the greatest chunk of the dataset representing more than half of the profiles collected. However, it is still sufficient to draw conclusions and answer the research questions of this study, as it captured deep insights into the recent years of the technology job market, which often periodically changes as new technologies emerge and others become obsolete. Third, since this analysis focused on the job market of the country for which the profiles were collected, Saudi Arabia, its findings may not necessarily generalize to a different country, as the job market and in-demand skills differ across countries. Lastly, different factors were not taken into account while performing the analysis, such as economic growth and joining tech boot camps tailored to fresh graduates. These factors could have some impact that is difficult to capture based solely on LinkedIn profiles.

Beyond the analysis of the collected dataset, state-of-the-art machine learning algorithms could be used to build predictive models. Such models could predict the best-suited computing major for newly enrolled students based on their interests and the characteristics of the roles that relate strongly to the given major. Therefore, students should take the major selection assessment based on interests and personality [22-23]. Then, the assessment results could be fed to the model to accurately propose a well-suited major. Deploying machine learning models could prevent potentially serious consequences on a student's journey and save significant resources that could result from selecting an unsuited field of study.

## VII. CONCLUSION AND FUTURE WORK

This study used public data from LinkedIn to analyze career opportunities for fresh computing graduates in Saudi Arabia. Various computing majors were compared, discovering interesting insights on how they may affect future career opportunities. This analysis shows certain relationships between computing majors and employed job positions after graduation. For instance, IS graduates are more often hired as business analysts. In addition, this study analyzed trends over the years for some job categories. The analysis showed that the demands for System Admin and Technical Support positions are declining, while the demands for Business Analysis and Artificial Intelligence & Data Science increase. The study also investigated the relationships between computing majors and professional certificates, discovering specific preferences across computing majors that are consistent with job position findings.

The future direction of this research will focus on using machine learning techniques to create tools that benefit computing students based on the data and findings of this research. These tools include recommending the major of study, the job preference, and the professional certifications to obtain. Similarly, machine learning models can be created based on LinkedIn data to benefit employees in developing a career path [24]. Another future research direction is to include graduate degrees, such as master's and Ph.D., and see how they impact future career opportunities.

## ACKNOWLEDGMENT

This research project was supported by a grant from the Researchers Supporting Project number RSPD2023R912, King Saud University, Riyadh, Saudi Arabia.